\documentclass[review=false]{jfp-epi}

\usepackage{latexsym}
\usepackage{graphicx}

\usepackage{tikz}
\usetikzlibrary{positioning}
\tikzset{mynode/.style={draw,circle,inner sep=2pt,outer sep=0pt}}

\usepackage {diagbox}

\received[Submitted]{2026-03-27}
\received[accepted]{0000-00-00}

\begin{document}

\title{On Representability of Multiple-Valued Functions by Linear Lambda Terms Typed with Second-order Polymorphic Type System}

\author{Satoshi Matsuoka}
\orcid{0000-0003-2126-2926}

\affiliation{%
  \institution{AIST}
  \city{Tsukuba}
  \country{Japan}
  \authoremail{s-matsuoka@aist.go.jp}
}

\begin{abstract}
  We show that any multiple-valued function can be represented by a linear lambda term typed in a second-order polymorphic type system, using two distinct styles.
  The first is a {\it circuit} style, which mimics combinational circuits in switching theory.
  The second is an {\it inductive} style, which follows a more traditional mathematical approach.
  We also discuss several optimizations for these representations.
  Furthermore, we present a case study that demonstrates the potential applications of our approach across various domains.
\end{abstract}

\maketitle
\section{Introduction}
In \cite{Mat16} besides presenting the main result of the paper, the author have also demonstrated that any two-valued function with an arbitrary number of arguments can be represented by a linear lambda term typed within a second-order polymorphic type system, 
where the  base type is 
$\mbox{\tt T}_{\rm 2} = \forall \mbox{\tt 'a}. \mbox{\tt ('a->'a)} \mbox{\tt ->} \mbox{\tt ('a->'a)} \mbox{\tt ->} \mbox{\tt ('a->'a)}$.
In this article, building on that result, we show that any multiple-valued function with an arbitrary number of arguments can be represented by a linear lambda term typed within a second-order polymorphic type system, 
where the base type is
\[ \mbox{\tt T}_{\rm r} = \forall \mbox{\tt 'a}. \overbrace{\mbox{\tt ('a->'a)} \mbox{\tt ->} \cdots \mbox{\tt ->} \mbox{\tt ('a->'a)}}^{r} \mbox{\tt ->} \mbox{\tt ('a->'a)} \]
where $r$ is the number of multiple-values instead of two-values $0$ and $1$. 

We construct such linear lambda terms in two distinctive styles: circuit style and inductive style.

In the circuit style one gives combinators mimicking combinational circuits in switching theory.
The distinctive feature of the style is use of {\it copy} combinators:
such a combinator has type $\mbox{\tt T}_{\rm r} \mbox{\tt ->} \mbox{\tt T}_{\rm r} {\tt *} \mbox{\tt T}_{\rm r}$
and duplicates another combinator given as its input, representing a value among $\{ 0, 1, \ldots, r-1 \}$.
The idea of the use of copy combinators was originally introduced in \cite{Mai04}.

On the other hand, in the inductive style, starting from combinators with
type $\mbox{\tt T}_{\rm r} \mbox{\tt ->} \mbox{\tt T}_{\rm r}$, representing one argument functions over $\{ 0, 1, \ldots, r-1 \}$,
one {\it inductively} construct combinators with type
$\overbrace{\mbox{\tt T}_{\rm r} \mbox{\tt ->} \cdots \mbox{\tt ->} \mbox{\tt T}_{\rm r}}^n \mbox{\tt ->} \mbox{\tt T}_{\rm r}$, 
representing $n$ argument functions over $\{ 0, 1, \ldots, r-1 \}$.
The distinctive feature of the style is that one can dispense with copy combinators.
Moreover, such an inductive method is a more traditional mathematical approach compared to the circuit approach.

In addition, we propose several optimization methods to compute the result of function application more efficiently (Section~\ref{secOpt}).

Multiple-valued logics have numerous applications in computer science and engineering:
for example, simulation of digital circuits (\cite{BS95}), logic synthesis (\cite{DWK16}), program analysis (\cite{HJS01,RSW04}),
model checking (\cite{BG99,CDEG03}), cryptography (\cite{HPS02}), access control (\cite{BDH07}), and machine learning (\cite{ZBBD97,PF98,LS03}).
Multiple valued logics can be implemented using multiple-valued functions.
The author believes that representing multi-valued functions by linear lambda terms has numerous applications, particularly in access control and machine learning.
As a case study we consider a majority function on Belnap bilattice (\cite{Bel77}) and optimize its representation in a circuit-style manner (Chapter~\ref{secCaseStudy}).

\section{Second-order polymorphic linear type system}
In this section we present a second-order polymorphic linear type system.
This system is the linear type system in \cite{Mat16} augmented with second-order quantifier.
Below we follow the standard convention with regard to variable binding and substitution.
For example see \cite{Bar84}.

Untyped terms we consider are as follows:
    \[
    {\tt t} ::= {\tt x} \, \, | \, \, {\tt t} \, \, {\tt t} \, \, | \, \, {\tt fn} \, \,  {\tt x=> t} \, \, | \, \, {\tt (t, t)} \, \, | \, \, {\tt let} \, \, {\tt val} \, \,  {\tt (x,y)=t} \, \, {\tt in} \, \,  {\tt t} \, \,  {\tt end}
    \]
Such a term is not necessarily a linear term which we want to discuss:
we only consider linear terms that are typed in the second-order polymorphic linear type system described in the following.

The definition of types is as follows:
\[ \mbox{\tt A} ::= \mbox{\tt 'a} \, \, \, \,
                 | \, \, \, \, \mbox{\tt A*A} \, \, \, \,
                 | \, \, \, \, \mbox{\tt A->A} \, \, \, \,
                 | \, \, \, \, \forall \mbox{\tt 'a}. \mbox{\tt A->A}
                 \]
where {\tt 'a} is an atomic type.

A typing environment denoted by $\Gamma$ or $\Delta$ is a list of pairs of a term variable and a type, ${\tt x1 : A1}, \ldots,  {\tt xn : An}$,
where $n \ge 0$. 

A typing judgement is a triple of a typing environment, a term ${\tt t}$ and a type ${\tt A}$ denoted by $\Gamma \vdash {\tt t : A}$.

The set of free type variables in type ${\tt A}$, denoted by ${\rm FTV}({\tt A})$, is defined inductively:
      \begin{eqnarray*}
        {\rm FTV}(\mbox{\tt 'a}) & = & \{ \mbox{\tt 'a} \}\\
        {\rm FTV}(\mbox{\tt A1*A2}) & = & {\rm FTV}(\mbox{\tt A1->A2}) \, \, = \, \,  {\rm FTV}(\mbox{\tt A1}) \cup {\rm FTV}(\mbox{\tt A2}) \\
        {\rm FTV}({\tt \forall \mbox{\tt 'a}. \mbox{\tt A}}) & = & {\rm FTV}(\mbox{\tt A}) \backslash \{ \mbox{\tt 'a} \}
    \end{eqnarray*}
The set of free type variables in typing environment $\Gamma$ is also defined:
    \begin{eqnarray*}
      {\rm FTV}()            & = & \emptyset \\
      {\rm FTV}({\tt x:A}, \Gamma) & = & {\rm FTV}({\tt A}) \cup {\rm FTV}(\Gamma) 
    \end{eqnarray*}
    
Our second-order linear type system is as follows:
\[
\frac{}{\mbox{\tt x:A} \vdash \mbox{\tt x:A}} \, \, ({\rm Id})
\quad \quad
\frac{\Gamma, \mbox{\tt x:A}, \mbox{\tt y:B}, \Delta \vdash \mbox{\tt t:C}}
     {\Gamma, \mbox{\tt y:B}, \mbox{\tt x:A}, \Delta \vdash \mbox{\tt t:C}}
     \, \, ({\rm Ex})
\]
\[
\frac{\mbox{\tt x:A}, \Gamma \vdash \mbox{\tt t:B}}
     {\Gamma \vdash \mbox{\tt fn x=>t:A->B}}
     \, \, ({\rm Abs})
\quad \quad 
\frac
{ \Gamma \vdash \mbox{\tt t} : \mbox{\tt A->B} \quad \Delta \vdash \mbox{\tt s:A} }
{ \Gamma, \Delta \vdash \mbox{\tt t} \, \mbox{\tt s} \mbox{\tt:B}}
\, \, ({\rm App})
\]
\[
\frac
{ \Gamma \vdash \mbox{\tt s:A} \quad \Delta \vdash \mbox{\tt t:B} }
{ \Gamma, \Delta \vdash \mbox{\tt (s,t):A*B}}
\, \, ({\rm Pair})
\quad \quad 
\frac
{ \Gamma \vdash \mbox{\tt s:A*B}
\quad \mbox{\tt x:A}, \mbox{\tt y:B}, \Delta \vdash \mbox{\tt t:C} }
{ \Gamma, \Delta \vdash \mbox{\tt let val (x,y)=s in t end:C} }
\, \, ({\rm Let})
\]
\[
\frac{\Gamma \vdash \mbox{\tt s:}\forall \mbox{\tt 'a.A}}
     {\Gamma \vdash \mbox{\tt s:A}[\mbox{\tt B}/\mbox{\tt 'a}]}
\, \, (\mbox{$\forall$-Inst}) 
\quad \quad
\frac
{ \Gamma \vdash \mbox{\tt s:A}}
{ \Gamma \vdash \mbox{\tt s:}\forall \mbox{\tt 'a.A}}
\, \, (\mbox{\tt 'a} \not\in {\rm FTV}(\Gamma))
\, \, (\mbox{$\forall$-Intro})     
\]

In the following sections, we consider only linear terms $s$ 
typed with the form
\[
\vdash \mbox{\tt s:A}
\]
in our type system.
Such a term is usually called a {\it combinator}.
We say a combinator $s$ is {\it monomorphically typed} if
there is a derivation $\vdash \mbox{\tt s:A}$ where
for each instantiation of rule $\forall$-Inst, its instantiated type {\tt B} is always a type variable.

In the following we construct combinators with thee kinds of types
(using auxiliary combinators, possibly with more higher-order types):
\begin{itemize}
\item base type: $\mbox{\tt T}_{\rm r}$,
\item arrow type: $\overbrace{\mbox{\tt T}_{\rm r} \mbox{\tt ->} \cdots \mbox{\tt ->}  \mbox{\tt T}_{\rm r}}^n \mbox{\tt ->} \mbox{\tt T}_{\rm r}$, and 
\item duplication type: $\mbox{\tt T}_{\rm r} \mbox{\tt ->} (\overbrace{\mbox{\tt T}_{\rm r}, \cdots, \mbox{\tt T}_{\rm r}}^n )$
\end{itemize}

In \cite{Mat16}, we discussed four reduction rules: \\
\ \ \ \ \ \ \ \ ($\beta_1$): {\tt (fn x=>t)s} \ \ $\Rightarrow_{\beta_1}$ \ \  {\tt t[s/x]} \\
\ \ \ \ \ \ \ \ ($\beta_2$): {\tt let val (x,y)=(u,v) in w end} \ \ $\Rightarrow_{\beta_2}$ \ \  {\tt w[u/x,v/y]} \\
\ \ \ \ \ \ \ \ ($\eta_1$): {\tt fn x=>(t x)} \ \ $\Rightarrow_{\eta_1}$ \ \  {\tt t} \\
\ \ \ \ \ \ \ \ ($\eta_2$): {\tt let val (x,y) = t in (x,y)} \ \ $\Rightarrow_{\eta_2}$ \ \ {\tt t} \\

and a commutative conversion.
In all combinators with three kinds of types that we construct, 
when a combinator with base type is applied to a combinator with arrow or duplication type,
the resulting combinator can be reduced to normal form using only $\beta_1$ and $\beta_2$.
So we do not have to consider $\eta_1$, $\eta_2$, or the commutative conversion.
\section{Representation of $r$-valued functions with an arbitrary number of arguments}
The notation
    \[
    {\tt fun} \, \, {\tt f} \, \, {\tt x}_1 \, \, \cdots \, \, {\tt x}_n \, \, {\tt =} \, \, {\tt g}
    \]
    denotes the term
    \[
    {\tt fn} \, \, {\tt x}_1 {\tt =>} \cdots {\tt =>} \, \, {\tt fn} \, \, {\tt x}_n {\tt =>} \, \, {\tt g}
    \]
    named ${\tt f}$.

The base type used in our representation is a polymorphic type, defined as follows:
\[ \mbox{\tt T}_{\rm r} = \forall \mbox{\tt 'a}. \overbrace{\mbox{\tt ('a->'a)} \mbox{\tt ->} \cdots \mbox{\tt ->} \mbox{\tt ('a->'a)}}^{r} \mbox{\tt ->} \mbox{\tt ('a->'a)}  \]
The number of normal terms of $\mbox{\tt T}_{\rm r}$ is $r !$.
But any choice of $r$ normal terms from $\mbox{\tt T}_{\rm r}$ may fail to yield a correct representation.
Although other choices are possible, we choose the following $r$ terms, which are considered as $r$ powers of a cyclic permutation with length $r$:
  \[
  \begin{array}{lcl}
    {\tt fun} \, \, {\tt v}\_{\tt 0} \, \, {\tt f}_{r-1} \, \, {\tt f}_{r-2} \cdots {\tt f}_{0} \, \, {\tt x}
    & = &
    {\tt f}_{0} ( {\tt f}_{1} ( \cdots ( {\tt f}_{r-1} \, \, {\tt x})  \cdots )) \\
    {\tt fun} \, \, {\tt v}\_{\tt 1} \, \, {\tt f}_{r-1} \, \, {\tt f}_{r-2} \cdots {\tt f}_{0} \, \, {\tt x}
    & = & {\tt f}_{1} ( {\tt f}_{2} ( \cdots ( {\tt f}_{0} \, \, {\tt x})  \cdots )) \\
    & & \cdots \\
    {\tt fun} \, \, {\tt v}\_{\tt r-1} \, \, {\tt f}_{r-1} \, \, {\tt f}_{r-2} \cdots {\tt f}_{0} \, \, {\tt x}
    & = & {\tt f}_{r-1} ( {\tt f}_{0} ( \cdots ( {\tt f}_{r-2} \, \, {\tt x})  \cdots ))
    \end{array}
  \]
  The standard $\mbox{\tt I}$ combinator plays an important role:
    \[
    {\tt fun} \, \, {\tt I} \, \, {\tt x} \, \, {\tt =} \, \, {\tt x}
    \]
 \subsection{Circuit style construction of typed linear lambda terms}
  The following term represents a one variable constant function:
  \[
    {\tt fun} \, \, {\tt const}\_{\tt i} \, \, {\tt h} \, \, {\tt f}_{r-1} \, \, {\tt f}_{r-2} \cdots {\tt f}_{0} \, \, {\tt x}
    = 
    {\tt f}_{{\rm i}} ( {\tt f}_{{\rm i}+1 ({\rm mod} \, {\rm r})} ( \cdots ( {\tt f}_{{\rm i}+{\rm r}-1 ({\rm mod} \, {\rm r})} \, \, ( {\tt h} \, \,  \overbrace{{\tt I} \cdots {\tt I}}^{r} \, \, {\tt x}) ) \cdots ))
    \]
    where $0 \le i \le r-1$.
    Note that ${\tt const}\_{\tt i}$ is monomorphically typed with $\mbox{\tt T}_{\rm r} \mbox{\tt ->} \mbox{\tt T}_{\rm r}$.

    Using these constant functions we can represent any one variable function $g : \{ 0, 1, \ldots, r-1 \} \to \{ 0, 1, \ldots, r-1 \}$:
    \[
    {\tt fun} \, \, {\tt g} \, \, {\tt h} 
    =
    {\tt h} \, \, {\tt const}\_{\tt g}_{r-1} \, \, {\tt const}\_{\tt g}_{r-2} \, \, \cdots \, \, {\tt const}\_{\tt g}_{0} \, \, {\tt v}\_{\tt 0}
    \]
    where ${\tt g}_{i}$ $\, \, (0 \le i \le r-1)$ denotes the function value $g(i)$.
    This term ${\tt g}$ can be typed with $\mbox{\tt T}_{\rm r} \mbox{\tt ->} \mbox{\tt T}_{\rm r}$.
    Note that in this typing,
    $\mbox{\tt T}_{\rm r}$ is substituted for the type variable {\tt 'a} in the type occurrence $\mbox{\tt T}_{\rm r}$ which the variable {\tt h} has.
    So it is not monomorphically typed.

    Thus for each $i, p \, (0 \le i, p \le r-1)$, the following one variable function  $C_i^p : \{ 0, 1, \ldots, r-1 \} \to \{ 0, 1, \ldots, r-1 \}$ can be represented:
    \[
    C_i^p(x) = \left\{
    \begin{array}{ll}
      p & (x = i) \\
      0   & (x \neq i)
    \end{array}
    \right.
    \]
    Such a function is called {\it literal} in the multiple-valued logic community.
    Let the corresponding combinator with type $\mbox{\tt T}_{\rm r} \mbox{\tt ->} \mbox{\tt T}_{\rm r}$ be ${\tt C}_i^p$.

    We note that the following combinator is monomorphically typed with $\mbox{\tt T}_{\rm r} \mbox{\tt ->} \mbox{\tt T}_{\rm r}$.
    \[
    {\tt fun} \, \, {\tt cyc}\_{\tt i} \, \, {\tt h} \, \, {\tt f}_{r-1} \, \, {\tt f}_{r-2} \cdots {\tt f}_{0} \, \, {\tt x}
    =
    {\tt h} \, \, {\tt f}_{{\rm i}} \, \, {\tt f}_{{\rm i} +1 ({\rm mod} \, {\rm r})} \cdots {\tt f}_{{\rm i}+{\rm r}-1 ({\rm mod} \, {\rm r})}\, \, {\tt v}\_{\tt 0} 
    \]
    which corresponds to the operation $x \mapsto x + i \, ({\rm mod} \, \, r)$, where $0 \le i \le r-1$.
    
    Next we consider the representation of two variable functions $M : {\{ 0, 1, \ldots, r-1 \} }^2 \to \{ 0, 1, \ldots, r-1 \}$,
    which can be considered as a square matrix with order $r$.
    For the preparation, we define a couple of auxiliary terms.

    We define an auxiliary term, representing a higher order function that
    receives function $F : \{ 0, \ldots, r-1 \} \to \{ 0, \ldots, r-1 \}$ and a value $v \in \{ 0, \ldots,  r-1 \}$ and
    always returns a fixed value $i \in \{ 0, \ldots, r-1 \}$:
    \[
    \begin{array}{ll}
    {\tt fun} \, \, {\tt const}\_{\tt f}\_{\tt i} \, \, {\tt F} \, \, {\tt h} \, \, {\tt f}_{r-1} \, \, {\tt f}_{r-2} \cdots {\tt f}_{0} \, \, {\tt x} 
    = \\
    \quad \quad \quad
          {\tt f}_{{\rm i}} ( {\tt f}_{{\rm i}+1 ({\rm mod} \, {\rm r})} ( \cdots ( {\tt f}_{{\rm i}+{\rm r}-1 ({\rm mod} \, {\rm r})} \, \, ({\tt h} \overbrace{{\tt I} \cdots {\tt I}}^{r} ( {\tt F} \, \, {\tt v}\_{\tt 0} \, \, \overbrace{{\tt I} \cdots {\tt I}}^{r} \, \,  {\tt x}) )) \cdots ))
    \end{array}
    \]    

    We define another auxiliary term, which
    represents the higher-order function that receives a function $F : \{ 0, \ldots, r-1 \} \to \{ 0, \ldots, r-1 \}$
    and a value $j \in \{ 0, \ldots, r-1 \}$ and returns the value $M(i, j) \in \{ 0, \ldots, r-1 \}$:
    \[
      {\tt fun} \, \, {\tt row}\_{\tt i} \, \, {\tt F} \, \, {\tt h} \, \,
      =
      \, \,
         {\tt h} \, \,
         {\tt const}\_{\tt f}\_{\tt M}_{i,r-1} {\tt const}\_{\tt f}\_{\tt M}_{i,r-2} \cdots {\tt const}\_{\tt f}\_{\tt M}_{i,0}
         \, \, {\tt I}         
         \,\, ({\tt F} \, \, {\tt v}\_{\tt 0})
    \]
    This term ${\tt row}\_{\tt i}$ can be typed with $\mbox{\tt (} \mbox{\tt T}_{\rm r} \mbox{\tt ->} \mbox{\tt T}_{\rm r} \mbox{\tt )} \mbox{\tt ->} \mbox{\tt (} \mbox{\tt T}_{\rm r} \mbox{\tt ->} \mbox{\tt T}_{\rm r} \mbox{\tt )}$.
    Note that in this typing,
    $\mbox{\tt (} \mbox{\tt (} \mbox{\tt T}_{\rm r} \mbox{\tt ->} \mbox{\tt T}_{\rm r} \mbox{\tt )} \mbox{\tt ->} \mbox{\tt (} \mbox{\tt T}_{\rm r} \mbox{\tt ->} \mbox{\tt T}_{\rm r} \mbox{\tt )} \mbox{\tt )} 
    \mbox{\tt ->}
    \mbox{\tt (} 
    \mbox{\tt (} \mbox{\tt T}_{\rm r} \mbox{\tt ->} \mbox{\tt T}_{\rm r} \mbox{\tt )} \mbox{\tt ->} \mbox{\tt (} \mbox{\tt T}_{\rm r} \mbox{\tt ->} \mbox{\tt T}_{\rm r} \mbox{\tt )}
    \mbox{\tt )}$
    is substituted for the type variable {\tt 'a} in the type occurrence $\mbox{\tt T}_{\rm r}$ which the bound variable {\tt h} holds.
    
    Then we can realize $M$ as follows:
    \[
      {\tt fun} \, \, {\tt M} \, \, {\tt h} \, \,
      =
      \, \,
         {\tt h} \, \, {\tt row}\_{\tt r-1} \, \, {\tt row}\_{\tt r-2} \, \, \cdots \, \, {\tt row}\_{\tt 0}
         \, \, {\tt I}          
         \]
    The combinator ${\tt M}$ can be typed with $\mbox{\tt T}_{\rm r} \mbox{\tt ->}\mbox{\tt T}_{\rm r} \mbox{\tt ->}\mbox{\tt T}_{\rm r}$.
         
    Thus we can represent the following $r$-valued conjunction $x \& y$ and disjunction $x \sqcup y$ as combinators:
    \begin{eqnarray*}
      x \& y & = & \min \{ x, y \} \\
      x \sqcup y & = & \max \{ x, y \}
    \end{eqnarray*}

    We can easily see that
    $n$ variable versions of the conjunction and disjunction can be realized as combinators.
    Let the generalized conjunction and disjunction with type
    $\overbrace{\mbox{\tt T}_{\rm r} \mbox{\tt ->} \cdots \mbox{\tt ->}  \mbox{\tt T}_{\rm r}}^n \mbox{\tt ->} \mbox{\tt T}_{\rm r}$
    be 
    $\&_n$ and $\sqcup_n$ respectively.

    In our type system, a duplication function like $x \mapsto (x, x)$ can not be represented without additional constraints.
    Nevertheless, when restricted to the base type, a duplication function can be expressed in a limited form.

    For this purpose, we need an auxiliary term to handle the function application of tuples:
         \[ {\tt fun} \, \,  {\tt tp}\_{\tt app} \, \,  {\tt h} \, \,  {\tt z} \, \,  =
         \, \, {\tt let} \, \, {\tt val} \, \, ({\tt f}, {\tt g}) = {\tt h} \, \, {\tt in} \, \,  {\tt let} \, \,  {\tt val} \, \, ({\tt x}, {\tt y}) = {\tt z} \, \, {\tt in} \, \, ({\tt f} \, \,  {\tt x}, {\tt g} \, \, {\tt y}) \, \, {\tt end} \, \,  {\tt end} \]
         Then we can represent the function returning $({\tt v}\_{\tt i}, {\tt v}\_{\tt i})$
         when receiving ${\tt v}\_{\tt i}$ for each $i \, (0 \le i \le r-1)$:
         \[
         \begin{array}{l}
           {\tt fun} \, \,  {\tt copy}_{\tt r} \, \, {\tt v} \, \, = \\
           \quad \quad {\tt let}  \, \, {\tt val} \, \, ({\tt x}, {\tt y}) = \\
           \quad \quad \quad {\tt v} \, \, ({\tt tp}\_{\tt app} \, \, ({\tt const}\_{\tt 0}, {\tt const}\_{\tt 0})) \cdots ({\tt tp}\_{\tt app}  \, \, ({\tt const}\_{\tt r-1}, {\tt const}\_{\tt r-1})) \\
           \quad \quad \quad \quad \quad \quad \quad \quad \quad \quad \quad \quad \quad \quad \quad \quad \quad \quad \quad \quad \quad \quad \quad \quad \quad \quad \quad \quad \quad ({\tt v}\_{\tt 0}, {\tt v}\_{\tt 0})\\
           \quad \quad \, \, {\tt in} \, \, ({\tt x}, {\tt y}) \, \, {\tt end}
         \end{array}
         \]
    The combinator ${\tt copy}_{\tt r}$ can be typed with $\mbox{\tt T}_{\rm r} \mbox{\tt ->} (\mbox{\tt T}_{\rm r}, \mbox{\tt T}_{\rm r})$.
    Moreover we can easily define 
    the generalized version of ${\tt copy}_{\tt r}$ with type
    $\mbox{\tt T}_{\rm r} \mbox{\tt ->} (\overbrace{\mbox{\tt T}_{\rm r}, \cdots, \mbox{\tt T}_{\rm r}}^n )$,
    which we call ${\tt copy}_{\tt r, n}$.

    Based on the preparations, and following the method described in \cite{Eps93} --- a generalization of the disjunctive normal form for Boolean functions ---
    we can represent any $n$-variable, $r$-valued function as a combinator with type
    $\overbrace{\mbox{\tt T}_{\rm r} \mbox{\tt ->} \cdots \mbox{\tt ->}  \mbox{\tt T}_{\rm r}}^n \mbox{\tt ->} \mbox{\tt T}_{\rm r}$.
                                   
    Suppose we are given an $n$ variable, $r$-valued function$ f : {\{ 0, 1, \ldots, r-1 \}}^n \to \{ 0, 1, \ldots, r-1 \}$.
    For each $(u_1, \ldots, u_{n}) \in  {\{ 0, 1, \ldots, r-1 \}}^n$, let $v = f(u_1, \ldots, u_n)$.
    We can then construct the following monomial term:
    \[
    \&_n ({\tt C}_{u_1}^{v} {\tt x}_1) \cdots ({\tt C}_{u_n}^{v} {\tt x}_n) 
    \]
    Next we combine these $r^n$ monomials with $\sqcup_{r^n}$.
    Then, by using ${\tt copy}_{{\tt r, r^n}}$, we can unify $r^n$ occurrences of the same variable in these monomials.
    We can easily see that the resulting combinator has type $\overbrace{\mbox{\tt T}_{\rm r} \mbox{\tt ->} \cdots \mbox{\tt ->}  \mbox{\tt T}_{\rm r}}^n \mbox{\tt ->} \mbox{\tt T}_{\rm r}$.
    Consequently we have achieved the representation of $r$-valued functions with an arbitrary number of arguments by polymorphically typed linear lambda terms.

    In our construction, we have used the linear order on $\{ 0, 1, \ldots, r-1 \}$.
    However, instead of a linear order, we may employ a general finite lattice (see \cite{Eps93}).
    In Section~\ref{secCaseStudy}, we consider the Belnap lattice, which consists of two four-element lattices.    
\subsection{Constant and projection functions}
Let ${\mbox{\tt U}}_{n, r}$ be $\overbrace{\mbox{\tt T}_{\rm r} \mbox{\tt ->} \cdots \mbox{\tt ->} \mbox{\tt T}_{\rm r}}^n \mbox{\tt ->} \mbox{\tt T}_{\rm r}$.    
We can easily construct a combinator with type ${\mbox{\tt U}}_{n, r}$
that represents an $n$ variable constant function ${\rm const}_{n,c}$.
That is, for some constant $c \in \{ 0, \ldots, r-1 \}$, and
for any $\mathbf{x} \in { \{ 0, \ldots, r-1 \} }^n$, we have ${\rm const}_{n,c}(\mathbf{x}) = c$:
\[
  {\tt fun} \, \, {\tt const}_{n, c} \, \, {\tt h}_1 \cdots {\tt h}_{n-1} \, \, {\tt h}_n
  = 
  ( {\tt h}_1 \overbrace{{\tt I} \cdots {\tt I}}^r) ( \cdots
  ( {\tt h}_{n-1} \overbrace{{\tt I} \cdots {\tt I}}^r)
  ( {\tt h}_n \overbrace{{\tt I} \cdots {\tt I}}^r \, \, {\tt c}) \cdots)
\]
where ${\tt c}$ is the term that represents $c$.

Similarly, we can construct a combinator with type ${\mbox{\tt U}}_{n, r}$
that represents an $n$ variable $i$-th projection ${\rm proj}_{n,i}$ on $n$ variables $(1 \le i \le n)$, i.e.,
${\rm proj}_{n,i}(x_1, \ldots, x_i, \ldots, x_n) = x_i$:
\[
\begin{array}{l}
  {\tt fun} \, \, {\tt proj}_{n, i} \, \, {\tt h}_1 \cdots {\tt h}_n
  = \\
  \quad \quad \quad \quad 
  ( {\tt h}_1 \overbrace{{\tt I} \cdots {\tt I}}^r) ( \cdots
  ( {\tt h}_{i-1} \overbrace{{\tt I} \cdots {\tt I}}^r) (
  ( {\tt h}_{i+1} \overbrace{{\tt I} \cdots {\tt I}}^r) (
  \cdots
  ( {\tt h}_n \overbrace{{\tt I} \cdots {\tt I}}^r \, \, {\tt h}_i) \cdots)) \cdots )
  \end{array}
\]
\subsection{Generalized multiple valued functions}
We can also easily construct a combinator with generalized type
\[ \overbrace{\mbox{\tt T}_{\rm r_1} \mbox{\tt ->} \cdots \mbox{\tt ->}  \mbox{\tt T}_{\rm r_k}}^n \mbox{\tt ->} \mbox{\tt T}_{\rm r_{k+1}} \]
that represents an arbitrary function
\[ f : \{ 0, \ldots, r_1-1 \} \times \cdots \times \{ 0, \ldots, r_k-1 \} \to \{ 0, \ldots, r_{k+1}-1 \}, \]
where $r_i \ge 2 \, (1 \le i \le k)$.
To show this, it suffices to construct a combinator with
\[ \mbox{\tt T}_{\rm r} \mbox{\tt ->} \mbox{\tt T}_{\rm r'} \]
that represents an arbitrary function
\[ g: \{ 0, \ldots, r-1 \} \to \{ 0, \ldots, r'-1 \}, \]
where $r, r' \ge 1$.
This can be achieved by modifying a combinator for a single-variable function:
replacing $\mbox{\tt T}_{\rm r}$ with $\mbox{\tt T}_{\rm r'}$ for the type variable {\tt 'a}.
\subsection{Inductive style construction of typed linear lambda terms}
In this section, we show that polymorphically typed linear lambda terms representing $n$ variable functions can be constructed inductively.
We fix $r \, (r \ge 1)$. 
We already know that one variable functions can be represented by terms with type $\mbox{\tt T}_{\rm r} \mbox{\tt ->} \mbox{\tt T}_{\rm r}$.
We can choose these terms ${\tt M}_{1, i} \, (1 \le i \le r^r)$ such that each ${\tt M}_{1, i}$ represents a distinct single variable function.
This forms the base step.

We now proceed to the the induction step.
Let ${\mbox{\tt U}}_{n, r}$ be $\overbrace{\mbox{\tt T}_{\rm r} \mbox{\tt ->} \cdots \mbox{\tt ->} \mbox{\tt T}_{\rm r}}^n \mbox{\tt ->} \mbox{\tt T}_{\rm r}$ as before.
Assume that we have a list of linear lambda terms representing $n$ variable functions
with type ${\mbox{\tt U}}_{n, r}$.
Let these terms be ${\tt M}_{n, i} \, (1 \le i \le r^{r^n})$. 
For each $i \, (1 \le i \le r^{r^n})$, from ${\tt M}_{n, i}$ we construct a term ${\tt M}_{n, i}^{\tt Fun}$ with type ${\mbox{\tt U}}_{n, r} \mbox{\tt ->} {\mbox{\tt U}}_{n, r} $
as follows:
\[
  {\tt fun} \, \, {\tt M}_{n, i}^{\tt Fun} \, \, {\tt F} \, \, {\tt h}_1 \cdots {\tt h}_n 
  \, \, \, \, = \, \, \, \, 
  ({\tt F} \, \, {\tt v}\_{\tt 0} \overbrace{{\tt I} \cdots {\tt I}}^r) \, \, ( {\tt M}_{n, i} \, \, {\tt h}_1 \cdots {\tt h}_n)
\]
Next, suppose that we are given a $n+1$-variable function $f: { \{ 0, \ldots, r-1 \} }^{n+1} \to { \{ 0, \ldots, r-1 \} }$.
Then for each $j \, (0 \le j \le r-1)$, let ${\tt M}_{n, k_j}$ be the term representing the $n$-variable function $\lambda x_1. \cdots \lambda x_{n}. f(j, x_1, \ldots x_{n})$.
Finally, we can construct a term ${\tt M}_{n+1, f}$ with type ${\mbox{\tt U}}_{n+1, r}$ representing $f$ as follows:
\[
  {\tt fun} \, \, {\tt M}_{n+1, f} \, \, {\tt h} \, \, \, \, = \, \, \, \, 
 {\tt h} \, \, {\tt M}_{n, k_{r-1}}^{\tt Fun} \cdots {\tt M}_{n, k_0}^{\tt Fun} \, \, {\tt const}_{n, c}
 \]
 We note that in the inductive-style construction, {\tt copy} combinators are not required, since we can directly derive a representation of arbitrary $n$ variable function.

  We observe that a hybrid approach combining the circuit style with the inductive style is feasible.
  For example, several linear lambda terms representing four-variable functions constructed inductively can be composed into
  a single four variable function in a circuit-style manner.
  Conversely, starting from several three-variable functions constructed in a circuit-style manner,
  one can compose
  a single four variable function in an inductive-style approach.
\section{Some optimizations}
\label{secOpt}
\subsection{Literal functions}
When we apply ${\tt const}\_{\tt i}$ to ${\tt v}\_{\tt j}$,
it requires $2r+1$ times applications of $\beta_1$-rule to reach its normal form.
In contrast, applying ${\tt I}$ to the same term requires only a single application.
Therefore, by replacing some occurrences of ${\tt const}\_{\tt i}$ by ${\tt I}$
in the representation of a one-variable function,
we can obtain a representation that is easier to reduce.

In particular, a literal $C_i^p$ is such a function, 
For example, if we consider the case where $r=5, p=4, i=3$, then we can the following optimized term:
  \[
    {\tt fun} \, \, {\tt C}_3^4 \, \, {\tt h} 
    =
    {\tt h} \, \, {\tt const}\_{\tt 0} \, \, {\tt const}\_{\tt 4} \, \, {\tt const}\_{\tt 0} \, \, {\tt I} \, \, {\tt I} \, \, {\tt v}\_{\tt 0}
  \]
\subsection{Two variable functions}
\subsubsection{Use of combinator {\tt I}}  
We can apply a similar optimization to two variable functions.
Suppose that we are given a two variable function specified by the following matrix:
\[
\begin{array}{|c||c|c|c|c|c|}
  \hline
    & 0 & 1 & 2 & 3 & 4 \\
  \hline\hline
  0 & 0 & 1 & 2 & 3 & 4 \\
  \hline
  1 & 0 & 1 & 2 & 4 & 3 \\
  \hline
  2 & 0 & 1 & 3 & 4 & 2 \\
  \hline
  3 & 0 & 2 & 3 & 4 & 1 \\
  \hline
  4 & 1 & 2 & 3 & 4 & 0 \\
  \hline
\end{array}
\]
When we implement each row using ${\tt row}\_{\tt i}$, then we need to include twenty-five ${\tt const}\_{\tt f}\_{\tt i}$ as auxiliary terms.
However, by considering transposition of the above matrix, we can derive the following optimized ${\tt row}\_{\tt i}$ terms:
\begin{eqnarray*}
      {\tt fun} \, \, {\tt row}\_{\tt 0} \, \, {\tt F} \, \, {\tt h} \, \,
      & = &
      \, \,
         {\tt h} \, \,
         {\tt const}\_{\tt f}\_{\tt 0} \, \, {\tt const}\_{\tt f}\_{\tt 1}  \, \, {\tt I} \, \, {\tt I} \, \, {\tt I} 
         \, \, {\tt I}         
         \,\, ({\tt F} \, \, {\tt v}\_{\tt 0})
         \\
      {\tt fun} \, \, {\tt row}\_{\tt 1} \, \, {\tt F} \, \, {\tt h} \, \,
      & = &
      \, \,
         {\tt h} \, \,
         {\tt const}\_{\tt f}\_{\tt 2} \, \, {\tt I} \, \, {\tt const}\_{\tt f}\_{\tt 1}  \, \, {\tt I} \, \, {\tt I} 
         \, \, {\tt I}         
         \,\, ({\tt F} \, \, {\tt v}\_{\tt 0})
         \\
      {\tt fun} \, \, {\tt row}\_{\tt 2} \, \, {\tt F} \, \, {\tt h} \, \,
      & = &
      \, \,
         {\tt h} \, \,
         {\tt const}\_{\tt f}\_{\tt 3} \, \, {\tt I} \, \, {\tt I} \, \, {\tt const}\_{\tt f}\_{\tt 2}  \, \, {\tt I} 
         \, \, {\tt I}         
         \,\, ({\tt F} \, \, {\tt v}\_{\tt 0})
         \\
      {\tt fun} \, \, {\tt row}\_{\tt 3} \, \, {\tt F} \, \, {\tt h} \, \,
      & = &
      \, \,
         {\tt h} \, \,
         {\tt const}\_{\tt f}\_{\tt 4} \, \, {\tt I} \, \, {\tt I} \, \, {\tt I} \, \, {\tt const}\_{\tt f}\_{\tt 3}  
         \, \, {\tt I}         
         \,\, ({\tt F} \, \, {\tt v}\_{\tt 0})
         \\
      {\tt fun} \, \, {\tt row}\_{\tt 3} \, \, {\tt F} \, \, {\tt h} \, \,
      & = &
      \, \,
         {\tt h} \, \,
         {\tt const}\_{\tt f}\_{\tt 4} \, \, {\tt const}\_{\tt f}\_{\tt 3}  \, \, {\tt const}\_{\tt f}\_{\tt 2} \, \, {\tt const}\_{\tt f}\_{\tt 1} \, \, {\tt const}\_{\tt f}\_{\tt 0}  
         \, \, {\tt I} 
         \,\, ({\tt F} \, \, {\tt v}\_{\tt 0})
\end{eqnarray*}
Then we need only thirteen ${\tt const}\_{\tt f}\_{\tt i}$ auxiliary terms.

But we can not apply such an optimization to the function specified by the following table, which is a Latin square:
\[
\begin{array}{|c||c|c|c|c|c|}
  \hline
    & 0 & 1 & 2 & 3 & 4 \\
  \hline\hline
  0 & 0 & 1 & 2 & 3 & 4 \\
  \hline
  1 & 1 & 3 & 0 & 4 & 2 \\
  \hline
  2 & 2 & 4 & 3 & 1 & 0 \\
  \hline
  3 & 3 & 0 & 4 & 2 & 1 \\
  \hline
  4 & 4 & 2 & 1 & 0 & 3 \\
  \hline
\end{array}
\]

\subsubsection{Optimization of modular addition}
Addition modulo $r$ is an important operation, as it is widely used in many areas.
The following matrix illustrates addition modulo $5$:
\[
\begin{array}{|c||c|c|c|c|c|}
  \hline
    & 0 & 1 & 2 & 3 & 4 \\
  \hline\hline
  0 & 0 & 1 & 2 & 3 & 4 \\
  \hline
  1 & 1 & 2 & 3 & 4 & 0 \\
  \hline
  2 & 2 & 3 & 4 & 0 & 1 \\
  \hline
  3 & 3 & 4 & 0 & 1 & 2 \\
  \hline
  4 & 4 & 0 & 1 & 2 & 3 \\
  \hline
\end{array}
\]
There is an optimization for modulo $r$ addition.
We introduce the following auxiliary term, which implements lifted cyclic operations:
  \[
    {\tt fun} \, \, {\tt cyc}\_{\tt f}\_{\tt i} \, \, {\tt F} \, \,{\tt h} \, \, {\tt f}_{r-1} \, \, {\tt f}_{r-2} \cdots {\tt f}_{0} \, \, {\tt x}
    =
    {\tt h} \, \, {\tt f}_{{\rm i}} \, \, {\tt f}_{{\rm i} +1 ({\rm mod} \, {\rm r})} \cdots {\tt f}_{{\rm i}+{\rm r}-1 ({\rm mod} \, {\rm r})}\, \, ( {\tt F} \, \, {\tt v}\_{\tt 0} \, \, \overbrace{{\tt I} \cdots {\tt I}}^{r} \, \,  {\tt x})
    \]
Using this, we can construct the desired term for addition modulo $r$:
    \[
      {\tt fun} \, \, {\tt add}\_{\tt mod} \, \, {\tt h} \, \,
      =
      \, \,
         {\tt h} \, \, {\tt cyc}\_{\tt f}\_{\tt r-1} \, \, {\tt cyc}\_{\tt f}\_{\tt r-2} \, \, \cdots \, \, {\tt cyc}\_{\tt f}\_{\tt 0}
         \, \, {\tt I}          
    \]
\section{Case study: representation of a majority function with more than two variables}
\label{secCaseStudy}
In the previous section we have discussed some optimizations of representation for multiple-valued functions with two arguments.
In this section, we turn to a more practical case study.
We examine how a four-argument majority function is represented on the Belnap bilattice in the circuit style and explore its optimizations.
\subsection{What is the Belnap bilattice?}
The Belnap bilattice \cite{Bel77} is a four-element set $\mathbf{4} = \{ \top, \bot, t, f \}$ equipped with two partial orders $\le_k$ and $\le_t$,
each forming a lattice structure. 

Informally, $t$ and $f$ represent truth values as in classical logic.
While $\bot$ indicates a lack of information, $\top$ indicates a kind of contradiction.
For example, suppose a reporter is trying to determine whether a claim is true or false.
One source states that the claim is true, while another states that it is false.
In this case, the claim should be assigned the value $\top$.

The two partial orders, $\le_k$ and $\le_t$, can be illustrated using Hasse diagrams as follows:
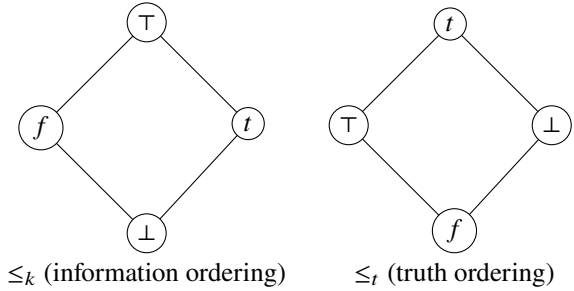
\begin{center}
    \begin{tikzpicture}

      \node [mynode] (top) at (0,0) {$\top$};
      \node [mynode,below left  = of top] (l)  {$f$};
      \node [mynode,below right = of top] (r) {$t$};
      \node [mynode,below right = of l,label=below: $\le_k$ (information ordering)] (bot) {$\bot$};
      \draw (top) -- (l) 
            (top) -- (r) 
            (l) -- (bot) 
            (r) -- (bot);

    \end{tikzpicture}
    \ \ \ \ 
    \begin{tikzpicture}

      \node [mynode] (top) at (0,0) {$t$};
      \node [mynode,below left  = of top] (l)  {$\top$};
      \node [mynode,below right = of top] (r) {$\bot$};
      \node [mynode,below right = of l, label=below: $\le_t$ (truth ordering)] (bot) {$f$};

      \draw (top) -- (l) 
            (top) -- (r) 
            (l) -- (bot) 
            (r) -- (bot);

    \end{tikzpicture}
\end{center}

The partial order $\le_k$ is called the {\it information ordering} and reflects the above informal explanation.
while $\top$ is the greatest element, $\bot$ is the least.
On the other hand, another partial order $\le_t$ is called the {\it truth ordering}:
neither $\top$ nor $\bot$ is considered true nor false.

As a lattice, the information ordering has the join and meet operators, usually denoted $\oplus$ and $\otimes$ respectively,
which is shown as a table as follows:
\[
\begin{array}{ccccc}
  \bigoplus:
  &
  \begin{array}{|c||c|c|c|c|}
    \hline
    \hbox{\diagbox{$x$}{$y$}}      & \bot & f    & t    & \top  \\
    \hline\hline
    \bot      & \bot & f    & t    & \top  \\
    \hline
    f         & f    & f    & \top & \top   \\
    \hline
    t         & t    & \top & t    & \top  \\
    \hline
    \top      & \top & \top & \top & \top   \\
    \hline
  \end{array}
  &
  &
  \bigotimes:
  &
  \begin{array}{|c||c|c|c|c|}
    \hline
    \hbox{\diagbox{$x$}{$y$}}      & \bot & f    & t    & \top  \\
    \hline\hline
    \bot      & \bot & \bot & \bot & \bot  \\
    \hline
    f         & \bot & f    & \bot & f    \\
    \hline
    t         & \bot & \bot & t    & t    \\
    \hline
    \top      & \bot & f    & t    & \top  \\
    \hline
  \end{array}
\end{array}
\]  
On the other hand, the join and meet operators of the truth ordering, 
usually denoted $\vee$ and $\wedge$ respectively,
which is shown as a table as follows:
\[
\begin{array}{ccccc}
  \bigvee:
  &
  \begin{array}{|c||c|c|c|c|}
    \hline
    \hbox{\diagbox{$x_1$}{$x_2$}}      & \bot & f    & t    & \top  \\
    \hline\hline
    \bot      & \bot & \bot & t    & t  \\
    \hline
    f         & \bot & f    & t    & \top   \\
    \hline
    t         & t    & t    & t    & t  \\
    \hline
    \top      & t    & \top & t    & \top     \\
    \hline
  \end{array}
  &
  &
  \bigwedge:
  \begin{array}{|c||c|c|c|c|}
    \hline
    \hbox{\diagbox{$x_1$}{$x_2$}}      & \bot & f    & t    & \top  \\
    \hline\hline
    \bot      & \bot & f    & \bot & f    \\
    \hline
    f         & f    & f    & f    & f    \\
    \hline
    t         & \bot & f    & t    & \top  \\
    \hline
    \top      & f    & f    & \top & \top   \\
    \hline
  \end{array}
\end{array}
\]

The Belnap bilattice has several applications in artificial intelligence and logic programming (\cite{Gin88,Fit91,AA98}).
It is also applied in the field of computer security, particularly, in access control.
In \cite{BH11}, they propose using the Belnap bilattice as the foundation of a language that describes access control policies.
Such a language specifies who can access various data objects on a network and under what conditions.
In this language, while $t$ and $f$ represent access granted and access denied respectively, $\top$ and $\bot$ mean undefined and conflicting access information respectively. 
\subsection{The specification of a four-argument majority function}
We consider a four-argument majority function $F(x_1, x_2, x_3, x_4)$ as defined informally as follows:
\begin{itemize}
\item
  If at least one of the four input variables is assigned to the value $\top$,
  then $F(x_1, x_2, x_3, x_4) = \top$:
  in other words, if at least one source reports that the claim is contradictory, we judge it as ``contradictory''.
\item
  Otherwise, 
\begin{itemize}
\item If at least three input variables are assigned the value $t$, then $F(x_1, x_2, x_3, x_4) = t$:
  this corresponds to the usual specification of a majority function.
\item If at least three input variables are assigned the value $\bot$, the value of $F(x_1, x_2, x_3, x_4) = \bot$:
  in this case we consider the information insufficient and judge ot as ``undefined''.
\item In the other cases, $F(x_1, x_2, x_3, x_4) = f$. 
\end{itemize}
\end{itemize}

To represent $F(x_1, x_2, x_3, x_4)$ as a combinator in our polymorphic linear type system, we adopt the following design decisions:
\begin{itemize}
\item The function $F(x_1, x_2, x_3, x_4)$ is decomposed into two-argument subfunctions. 
\item The decomposition is based on the information ordering $\le_k$:
  we use $\oplus$ and $\otimes$ in place of $\max$ and $\min$ respectively.
\end{itemize}
Under the design decision, the function $F(x_1, x_2, x_3, x_4)$ is decomposed as
\begin{equation}
  \label{eq-DecomposedFunction-1}
\bigoplus_{i=1}^{7} f_i(x_1, x_2) \otimes g_i(x_3, x_4)
\end{equation}
where $f_i$ and $g_i$ $\, (1 \le i \le 7)$ are defined according to the values assigned to $x_1$ and $x_2$. 
\begin{enumerate}
\item [(1)] The case where $x_1 = \top$ and $x_2 = \top$:
  \[
  \begin{array}{ccccc}
    f_1:
    &
    \begin{array}{|c||c|c|c|c|}
      \hline
      \hbox{\diagbox{$x_1$}{$x_2$}}      & \bot & f    & t    & \top  \\
      \hline\hline
      \bot      & \bot & \bot & \bot & \top \\
      \hline
      f         & \bot & \bot & \bot & \top  \\
      \hline
      t         & \bot & \bot & \bot & \top  \\
      \hline
      \top      & \top & \top & \top & \top  \\
      \hline
    \end{array}
    &
    &
    g_1:
    &
    \begin{array}{|c||c|c|c|c|}
      \hline
      \hbox{\diagbox{$x_3$}{$x_4$}}      & \bot & f    & t    & \top  \\
      \hline\hline
      \bot      & \top & \top & \top & \top  \\
      \hline
      f         & \top & \top & \top & \top  \\
      \hline
      t         & \top & \top & \top & \top  \\
      \hline
      \top      & \top & \top & \top & \top  \\ 
      \hline
    \end{array}
  \end{array}
\]  

\item [(2)]The case where $x_1 = x_2 = t$:
  \[
  \begin{array}{ccccc}
    f_2:
    &
    \begin{array}{|c||c|c|c|c|}
      \hline
      \hbox{\diagbox{$x_1$}{$x_2$}}      & \bot & f    & t    & \top  \\
      \hline\hline
      \bot      & \bot & \bot & \bot & \bot  \\
      \hline
      f         & \bot & \bot & \bot & \bot   \\
      \hline
      t         & \bot & \bot & \top & \bot   \\
      \hline
      \top      & \bot & \bot & \bot & \bot    \\
      \hline
    \end{array}
    &
    &
    g_2:
    \begin{array}{|c||c|c|c|c|}
      \hline
      \hbox{\diagbox{$x_3$}{$x_4$}}      & \bot & f    & t    & \top  \\
      \hline\hline
      \bot      & f    & f    & t    & \top  \\
      \hline
      f         & f    & f    & t    & \top   \\
      \hline
      t         & t    & t    & t    & \top  \\
      \hline
      \top      & \top & \top & \top & \top   \\
      \hline
    \end{array}
  \end{array}
\]  

\item [(3)]
  The case where $x_1 = t$ and $x_2 = \bot$,  or $x_1 = \bot$ and $x_2 = t$:
  \[
  \begin{array}{ccccc}
    f_3:
    &
    \begin{array}{|c||c|c|c|c|}
      \hline
      \hbox{\diagbox{$x_1$}{$x_2$}}      & \bot & f    & t    & \top  \\
      \hline\hline
      \bot      & \bot & \bot & \top & \bot  \\
      \hline
      f         & \bot & \bot & \bot & \bot  \\
      \hline
      t         & \top & \bot & \bot & \bot \\
      \hline
      \top      & \bot & \bot & \bot &  \bot  \\
      \hline
    \end{array}
    &
    &
    g_3:
    &
    \begin{array}{|c||c|c|c|c|}
      \hline
      \hbox{\diagbox{$x_3$}{$x_4$}}      & \bot & f    & t    & \top  \\
      \hline\hline
      \bot      & \bot & f    & f    & \top  \\
      \hline
      f         & f    & f    & f    & \top   \\
      \hline
      t         & f    & f    & t    & \top  \\
      \hline
      \top      & \top & \top & \top & \top  \\
      \hline
    \end{array}
  \end{array}
\]  
\item [(4)]
  The case where $x_1 = f$ and $x_2 = \bot$,  or $x_1 = \bot$ and $x_2 = f$:
  \[
  \begin{array}{ccccc}
    f_4:
    &
    \begin{array}{|c||c|c|c|c|}
      \hline
      \hbox{\diagbox{$x_1$}{$x_2$}}      & \bot & f    & t    & \top  \\
      \hline\hline
      \bot      & \bot & \top & \bot & \bot  \\
      \hline
      f         & \top & \bot & \bot & \bot  \\
      \hline
      t         & \bot & \bot & \bot & \bot  \\
      \hline
      \top      & \bot & \bot & \bot & \bot   \\
      \hline
    \end{array}
    &
    &
    g_4:
    &
    \begin{array}{|c||c|c|c|c|}
      \hline
      \hbox{\diagbox{$x_3$}{$x_4$}}      & \bot & f    & t    & \top  \\
      \hline\hline
      \bot      & \bot  & f    & f    & \top  \\
      \hline
      f         & f    & f    & f    & \top   \\
      \hline
      t         & f    & f    & f    & \top  \\
      \hline
      \top      & \top & \top & \top & \top   \\
      \hline
    \end{array}
  \end{array}
\]  

\item [(5)]
  The case where $x_1 = x_2 = f$:
  \[
  \begin{array}{ccccc}
    f_5:
    &
    \begin{array}{|c||c|c|c|c|}
      \hline
      \hbox{\diagbox{$x_1$}{$x_2$}}      & \bot & f    & t    & \top  \\
      \hline\hline
      \bot      & \bot & \bot & \bot & \bot  \\
      \hline
      f         & \bot & \top & \bot & \bot  \\
      \hline
      t         & \bot & \bot & \bot & \bot  \\
      \hline
      \top      & \bot & \bot & \bot & \bot   \\
      \hline
    \end{array}
    &
    &
    g_5:
    &
    \begin{array}{|c||c|c|c|c|}
      \hline
      \hbox{\diagbox{$x_3$}{$x_4$}}      & \bot & f    & t    & \top  \\
      \hline\hline
      \bot      & f    & f    & f    & \top  \\
      \hline
      f         & f    & f    & f    & \top   \\
      \hline
      t         & f    & f    & f    & \top  \\
      \hline
      \top      & \top & \top & \top & \top   \\
      \hline
    \end{array}
  \end{array}
\]  

\item [(6)]
  The case where $x_1 = t$ and $x_2 = f$, or $x_1 = f$ and $x_2 = t$:
  \[
  \begin{array}{ccccc}
    f_6:
    &
    \begin{array}{|c||c|c|c|c|}
      \hline
      \hbox{\diagbox{$x_1$}{$x_2$}}      & \bot & f    & t    & \top  \\
      \hline\hline
      \bot      & \bot & \bot & \bot & \bot  \\
      \hline
      f         & \bot & \bot & \top & \bot  \\
      \hline
      t         & \bot & \top & \bot & \bot  \\
      \hline
      \top      & \bot & \bot & \bot & \bot     \\
      \hline
    \end{array}
    &
    &
    g_6:
    &
    \begin{array}{|c||c|c|c|c|}
      \hline
      \hbox{\diagbox{$x_3$}{$x_4$}}      & \bot & f    & t    & \top  \\
      \hline\hline
      \bot      & f    & f    & f    & \top  \\
      \hline
      f         & f    & f    & f    & \top  \\
      \hline
      t         & f    & f    & t    & \top  \\
      \hline
      \top      & \top & \top & \top & \top  \\
      \hline
    \end{array}
  \end{array}
\]  

\item [(7)]
  The case where $x_1 = x_2 = \bot$:
  \[
  \begin{array}{ccccc}
    f_7:
    &
\begin{array}{|c||c|c|c|c|}
  \hline
 \hbox{\diagbox{$x_1$}{$x_2$}}      & \bot & f    & t    & \top  \\
  \hline\hline
                          \bot      & \top & \bot & \bot & \bot  \\
  \hline
                           f         & \bot & \bot & \bot & \bot  \\
  \hline
                           t         & \bot & \bot & \bot & \bot  \\
  \hline
                          \top      & \bot & \bot & \bot & \bot  \\
  \hline
\end{array}
   &
   &
   g_7:
   &
\begin{array}{|c||c|c|c|c|}
  \hline
  \hbox{\diagbox{$x_3$}{$x_4$}}      & \bot & f    & t    & \top  \\
  \hline\hline
                           \bot      & \bot & \bot & \bot & \top  \\
  \hline
                           f         & \bot & f    & f    & \top  \\
  \hline
                           t         & \bot & f    & f    & \top  \\
  \hline
                           \top      & \top & \top & \top & \top  \\
  \hline
\end{array}
  \end{array}
\]  
\end{enumerate}
\subsection{A method for reducing the number of subfunctions}
We can apply certain optimizations to the decomposed function (\ref{eq-DecomposedFunction-1}).
In this section, we examine an optimization for reducing the number of two-argument subfunctions in the decomposed function.

For the decomposition function
\[
\bigoplus_{i=1}^{7} f_i(x_1, x_2) \otimes g_i(x_3, x_4) \, , 
\]
we observe that $g_3(x_3, x_4)$ and $g_4(x_3, x_4)$ are very similar:
they differ only in that $g_3(t, t) = t$ whereas $g_4(t, t) = f$.
To capture this difference, 
we define $\theta_1, \theta_2 : \{ t, f, \top, \bot \} \to \{ t, f, \top, \bot \}$ as follows:
\[
\theta_1(x) = x 
\quad \quad {\rm and} \quad \quad
\theta_2(x) =
\left\{
\begin{array}{ll}
  f & (x = t)     \\
  x & (x \neq t)
\end{array}
\right.
\]
Then there is a function $g_0 : {\{ t, f, \top, \bot \}}^2 \to \{ t, f, \top, \bot \}$ (in this case $g_0 = g_3$) such that
$g_3(x_3, x_4) = \theta_1 \circ g_0(x_3, x_4)$ and $g_4(x_3, x_4) = \theta_2 \circ g_0(x_3, x_4)$.

Furthermore, we note that 
\[
{\rm Im}(f_3) = {\rm Im}(f_4) = \{ \top, \bot \} \quad \quad {\rm and} \quad \quad
f_3^{-1}(\top) \cap f_4^{-1}(\top) = \emptyset, 
\]
where ${\rm Im}(g)$ denotes the image of the function $g$.
We define $f_0 : {\{ t, f, \top, \bot \}}^2 \to \{ t, f, \top, \bot \}$ as
\[
f_0(x_1, x_2) =
\left\{
\begin{array}{ll}
  f & (f_3(x_1, x_2) = \top) \\
  t & (f_4(x_1, x_2) = \top) \\
  \bot & ({\rm Otherwise})
\end{array}
\right.
\]
Next, we define $h_0 : {\{ t, f, \top, \bot \}}^2 \to \{ t, f, \top, \bot \}$ as
\[
\begin{array}{|c||c|c|c|c|}
  \hline
  \hbox{\diagbox{$x$}{$y$}}      & \bot & f    & t    & \top  \\
  \hline\hline
                           \bot      & \bot & \bot & \bot & \bot  \\
  \hline
                           f         & \theta_1(\bot) & \theta_1(f)    & \theta_1(t)    & \theta_1(\top)   \\
  \hline
                           t         & \theta_2(\bot)  & \theta_2(f)    & \theta_2(f)    & \theta_2(\top)  \\
  \hline
                           \top      & \top & \top & \top & \top   \\
  \hline
\end{array}
\]
Then  we have
\[
h_0(f_0(x_1, x_2), g_0(x_3, x_4)) = f_3(x_1, x_2) \otimes g_3(x_3, x_4) \oplus f_4(x_1, x_2) \otimes g_4(x_3, x_4) 
\]
We have successfully reduced the number of two-argument subfunctions by four. 
It should be straightforward to generalize this method. 

We remark that the above method can also be applied to 
\[
f_1(x_1, x_2) \otimes g_1(x_3, x_3) \oplus f_2(x_1, x_2) \otimes g_i(x_3, x_3) \, (2 \le i \le 7)
\]
and similarly to
\[
f_5(x_1, x_2) \otimes g_5(x_3, x_3) \oplus f_6(x_1, x_2) \otimes g_6(x_3, x_3) 
\]
or
\[
f_5(x_1, x_2) \otimes g_5(x_3, x_3) \oplus f_7(x_1, x_2) \otimes g_7(x_3, x_3) 
\]
or 
\[
f_6(x_1, x_2) \otimes g_6(x_3, x_3) \oplus f_7(x_1, x_2) \otimes g_7(x_3, x_3) 
\]
\subsection{Simplification of $\otimes$ and $\oplus$ using don't cares}
We note that the images of subfunctions $f_1$ and $g_1$ are $\{ \bot, \top \}$ and $\{ \top \}$ respectively.
Then in $f_1(x_1, x_2) \otimes g_1(x_3, x_4)$, for example, the pair $t \otimes f$ never occurs.
When we define a degenerate form of $\otimes$, denoted by $\otimes' : {\{ t, f, \top, \bot \}}^2 \to \{ \top, \bot \}$ as
\[
\begin{array}{cc}
  \otimes':
  &
\begin{array}{|c||c|c|c|c|}
  \hline
  \hbox{\diagbox{$x$}{$y$}}      & \bot & f    & t    & \top  \\
  \hline\hline
                           \bot  & \bot & \bot & \bot & \top \\
  \hline
                           f     & \top & \top & \top & \top \\
  \hline
                           t     & \top & \top & \top & \top  \\
  \hline
                           \top  & \top & \top & \top & \top  \\
  \hline
\end{array}
\end{array}
\]
we still have $f_1 \otimes' g_1 = f_1 \otimes g_1$. 
A similar optimization can also be applied to $f_i(x_1, x_2) \otimes g_i(x_3, x_4)$ $\, (2 \le i \le 6)$.
Moreover, 
a similar optimization can also be applied to 
the four occurrences of $\oplus$ in 
$\bigoplus_{i=1}^{7} f_i(x_1, x_2) \otimes g_i(x_3, x_4)$.
\section{Concluding remarks}
We have demonstrated that our linear lambda type system is sufficiently expressive to represent multiple-valued functions
using by not only through the existing circuit-based method, but also via an inductive method.

Moreover, we have examined several optimization methods and presented a case study.

In future work, we plan to:
\begin{itemize}
\item Comparison of three styles:
  We have discussed three styles --- circuit style, inductive style, and hybrid style.
  From a practical point view, a comparison of the efficiency of the three styles should be conducted.

\item Efficient reduction of linear lambda terms:
  Linear lambda terms have a distinctive structure: each bound variable is used {\it exactly once}.
  Therefore compared to the conventional $\beta$-reduction in the simply typed $\lambda$-calculus,
  our reduction rules, $\beta_1$ and $\beta_2$, are extremely simple.
  It appears that studies on hardware architectures and software implementation techniques to exploit this distinctive feature are still lacking.

  In addition, unlike PCF \cite{Plo77}, our linear lambda calculus does not include any if-then-else expressions, despite its rich expressiveness.
  An if-then-else expression is typically compiled into machine-level instructions involving branching.
  It is well known that branching is a bottleneck in microprocessors with pipelined architectures \cite{HP17}.
  It is also well known that conditional branching can be a source of information leakage \cite{Den82}.
  This feature may have potential new applications.  
\item Machine learning applications:
  Recent advances in AI techniques based on neural networks have been remarkable.
  However, as demonstrated in \cite{LLLRY19}, traditional logic-based machine learning approaches remain useful in several domains.
  Our representation of multiple-valued functions using linear lambda terms appears to offer significant benefits as building blocks for the implementation of such logic-based machine learning methods.
\end{itemize}

\bibliographystyle{ACM-Reference-Format}
\bibliography{generic} 

\label{lastpage01}

\end{document}